\documentclass[doublecol]{epl2} 

\usepackage{graphicx}
\usepackage{bm}
\usepackage{color}
\usepackage{amsmath,amsfonts,amssymb} 
\usepackage{verbatim}
\usepackage[normalem]{ulem}

\renewcommand{\revision}{}

\definecolor{cadmiumgreen}{rgb}{0.0, 0.42, 0.24}

\title{Polynomial complexity despite the fermionic sign
}

\author{R. Rossi
\inst{1}
\and
N. Prokof'ev
\inst{2,3}
\and
B. Svistunov
\inst{2,3,4,5,1}
\and
K. Van Houcke
\inst{1}
\and
F. Werner
\inst{5}
}

\shortauthor{R. Rossi \etal}

\institute{                    
  \inst{1} Laboratoire de Physique Statistique, Ecole Normale Sup\'erieure, UPMC, Universit\'e Paris Diderot, CNRS,\\ 24 rue Lhomond, 75005 Paris, France
\\
\inst{2}
Department of Physics, University of Massachusetts, Amherst, MA 01003, USA
\\
\inst{3}
National Research Center ``Kurchatov Institute,'' 123182 Moscow, Russia
\\
\inst{4}
Wilczek Quantum Center, Zhejiang University of Technology, Hangzhou 310014, China
\\
\inst{5}
Laboratoire Kastler Brossel, Ecole Normale Sup\'erieure, UPMC, CNRS, Coll\`ege de France, \\24 rue Lhomond, 75005 Paris, France
}

\abstract{
It is commonly believed that in 
\revision{unbiased}
quantum Monte Carlo approaches to fermionic many-body problems,
the infamous sign problem generically implies 
prohibitively large
computational times 
for obtaining thermodynamic-limit quantities. 
We point out
that for convergent
Feynman diagrammatic series evaluated
with the Monte Carlo algorithm of [Rossi, arXiv:1612.05184],
the computational time increases only polynomially with the inverse error 
on thermodynamic-limit quantities.
}

\pacs{02.70.Ss}{Quantum Monte Carlo methods}
\pacs{71.10.Fd}{Lattice fermion models (Hubbard model, etc.)}

\begin{document}
\maketitle

The notion of fermion sign problem (FSP) was originally formulated in the context
of auxiliary-field, path-integral and diffusion quantum Monte Carlo (QMC)
methods \cite{Loh,CeperleyPIQMC,CeperleyHouches,AssadChapter}.
There, it was observed that the computational time required for calculating
properties of the fermionic system to a given accuracy
scales exponentially with the system volume.
Later, the notion of FSP was implicitly extended to an arbitrary QMC
approach dealing with interacting fermions, referring to 
the stochastic sampling of a non-sign-definite
quantity with a near cancellation between positive and negative contributions.
Sign-free Monte Carlo (MC)
algorithms were emerging only as exceptions confirming the rule:
In each such case, the absence of FSP was due to some special property of
the simulated model (see, {\it e.g.}, \cite{WieseMeron,SachdevSignFree,Wei,AletSignFreeFrustratedSpins,HoneckerSignFreeFrustratedSpins,troyer_topol_origin,capponi_honeycomb}
and Refs. therein). Nowadays the FSP is generally perceived
as one of the most important unsolved problems in the field of numerical
studies of interacting fermionic systems in dimensions
$d>1$.\footnote{Frustrated spin and frustrated bosonic (with restricted on-site Hilbert space)
lattice models can be mapped to a system of interacting fermions and thus are part of the
present discussion.}

The main message of this letter follows from a simple observation.
Suppose some quantity $Q$ is computed from a limit $Q = \lim_{n \to \infty} Q_{n} $, with an
exponentially fast convergence, $\vert Q-Q_{n} \vert \sim e^{-\# n}$ (where $\#$ denotes some positive constant).
Then in order to compute $Q$ up to an error $| Q-Q_{n}|=\epsilon$ it is sufficient to take $n \sim \ln \epsilon^{-1}$.
Hence, even if the computational time increases exponentially as a function of $n$, $t \sim e^{\# n}$, the increase of $t$ as a function
of $\epsilon^{-1}$ is only polynomial, $\ln t \sim \ln \epsilon^{-1}$. 

This observation applies to the simulation of interacting fermions by the algorithm introduced in Ref.~\cite{Rossi}, denoted hereafter by the acronym CDet, for Connected Determinant
Diagrammatic MC. This algorithm works directly in the thermodynamic limit since it evaluates the series of {\it connected} Feynman diagrams. It exploits two advantages of the fermionic sign:
First, for fermions on a lattice at finite temperature, the series has a finite radius of convergence, so that the convergence as a function of diagram order $n$ is exponential;
second, a factorial number of connected Feynman diagrams 
can be evaluated in exponential time using determinants (and a recursive formula).

In general,
sign-alternation of observables simulated by MC methods
is neither sufficient nor necessary to state that the problem is intractable. One should rather focus
on what we will call the ``computational complexity problem'' (CCP) instead of the FSP.
The key question is the one that is most relevant practically\footnote{We closely follow ideas expressed by D. Ceperley at the Meeting of the {\it Simons collaboration on the many-electron problem}, New~York, Nov. 19-20, 2015.}:
How easily can one indefinitely 
increase the accuracy of the computed
thermodynamic-limit
 answer?
This leads to
the following definition of the CCP that can be applied to any numerical scheme.
Let $Q$  be the intensive quantity of interest in the thermodynamic limit. 
{\it  A numerical scheme is said to have CCP if the computational time $t$
required to obtain $Q$ with an error $\epsilon$ diverges faster than any
polynomial function of $\epsilon^{-1} \to \infty$.
The CCP is considered to be solved if}
\begin{equation}
t(\epsilon)\; = \; O(\epsilon^{-\alpha}).
\end{equation}
\revision{Note that we consider unbiased methods, i.e., $\epsilon\to0$ is the difference between computed value and {\it exact} value.}

In what follows, we 
show in some detail that CDet solves the CCP,
\revision{at least at finite temperature and small enough interaction}.
In parallel, we also discuss the conventional diagrammatic Monte Carlo approach (hereafter denoted by DiagMC) in which the sum over diagram topologies is done stochastically~\cite{diagmchub}.
We then 
show that
the conventional FSP leads to a CCP for
path-integral and auxiliary-field QMC.

In quantum Monte Carlo, one typically generates configurations $\mathcal{C}$ according to
  a conveniently chosen unnormalised probability distribution $P(\mathcal{C})$  that is positive. Any sign alternation is taken
  into account when collecting statistics, and is absorbed into the quantity   $A(\mathcal{C})$ that is being measured.
The average with respect to $P$,
  \begin{equation}
  \langle  A  \rangle_{P} = \frac{\sum_{\mathcal{C}}   P(\mathcal{C}) A(\mathcal{C})   }{\sum_{\mathcal{C}}  P(\mathcal{C}) } \; ,
  \label{eq:avA}
  \end{equation}
  is estimated through
  \begin{equation}
  \frac{1}{N_{MC}} \sum_{i=1}^{N_{MC}} A(\mathcal{C}_i) \;,
  \label{eq:avAMC}
  \end{equation}
  with $N_{MC}$ the number of MC measurements and $\mathcal{C}_i$ the configuration at the $i$-th measurement.
  By the central limit theorem, the $1\sigma$ statistical error on (\ref{eq:avAMC}) is given by
  \begin{equation}
  \epsilon_{{\rm stat}} = \sigma_A   \sqrt{\frac{2~\tau_{\rm auto}+1}{N_{MC}}} \; ,
  \label{eq:epsstat}
  \end{equation}
with $\tau_{\rm auto}$ the integrated autocorrelation time and $\sigma_A^{\phantom{.}2} = \langle A^2 \rangle_P - \langle A \rangle_P^{\phantom{.}2}$ the variance on individual measurements.

We now specialize to CDet and DiagMC.
We consider the computation of an observable ({\it e.g.} density or double occupancy).
For convenience,
we make two simplifications regarding the Monte Carlo algorithm.
We expect that this does not change the final CCP scaling.
The first simplification is that a separate simulation is performed for each order, while the normalisation factors $z_n$ (see below) are known.
The second simplification is that
in DiagMC, 
rather than sampling
the self-energy diagrams and then obtaining
observables from the Dyson equation, 
we consider 
here sampling
the diagrams for the observable (including one-particle reducible diagrams), 
so that external variables are simply fixed (to zero in space and imaginary-time representation).
A DiagMC configuration
is then defined by
  a Feynman diagram topology
together with values of the
internal variables $X$.
In CDet a configuration  is defined only by the internal variables $X$ (the space and imaginary-time coordinates of the interaction vertices),
while
the weight of a configuration is given by the sum over all possible connected diagram topologies connecting the internal and  external vertices.

Let us denote the contribution of a diagram of topology $\mathcal{T}$
for fixed internal
variables $X$ 
by $\mathcal{D}(\mathcal{T},X)$.
Let $a_n$ be the sum of all Feynman diagrams of order $n$:
\begin{eqnarray}
a_n  =   \int dX \sum_{\mathcal{T} \in \mathcal{S}_n} \mathcal{D}(\mathcal{T},X)  \; ,
\end{eqnarray}
with $\mathcal{S}_n$ the set of all diagram topologies at order $n$.
This can be rewritten in the form of Eq.~(\ref{eq:avA}):
\begin{equation}
a_n   =  \langle  A_n  \rangle_{P_n}  \; ,
\end{equation}
with the unnormalised distribution to be sampled chosen to be
\begin{equation}
 \left\{
                \begin{array}{ll}
      \displaystyle
        P_n(\mathcal{T},X) =    \left|   \mathcal{D}(\mathcal{T},X)     \right|\quad &   {\rm (DiagMC)} \\
\displaystyle
              P_n(X) =
                  \left|  \sum_{\mathcal{T}\in \mathcal{S}_n}  \mathcal{D}(\mathcal{T},X)     \right|  \quad &  {\rm (CDet)} 
                \end{array}
              \right.
              \end{equation}
              and
\begin{equation}
 A_n =\left\{
                \begin{array}{ll}
\displaystyle
                 z_n ~ {\rm sign}[\mathcal{D}(\mathcal{T},X)]  \quad &   {\rm (DiagMC)} \\
\displaystyle
                  z_n ~ {\rm sign}\left[\sum_{\mathcal{T}\in \mathcal{S}_n}  \mathcal{D}(\mathcal{T},X)\right]   \quad &  {\rm (CDet)} 
                \end{array}
              \right.
              \end{equation}
with the normalization factors
\begin{equation}
 z_n =\left\{
                \begin{array}{ll}
\displaystyle
                  \int dX \sum_{\mathcal{T}\in \mathcal{S}_n} |\mathcal{D}(\mathcal{T}, X)| \quad &   {\rm (DiagMC)} \\
\displaystyle
                  \int dX
\left|\sum_{\mathcal{T}\in \mathcal{S}_n} \mathcal{D}(\mathcal{T}, X) \right|
  \quad &  {\rm (CDet)} \; .
                \end{array}
              \right.
              \end{equation}
So,  in DiagMC the diagrams are sampled according to the distribution $  \left|   \mathcal{D}(\mathcal{T},X)     \right|$,
while in CDet diagrams are grouped together via determinants and in the MC part of the algorithm one samples $X$ according to the distribution
$\left|  \sum_{\mathcal{T}\in \mathcal{S}_n}  \mathcal{D}(\mathcal{T},X)     \right|$.
In what follows we neglect the statistical error on the normalisation factors
$z_n$ since they are obtained by sampling a sign-positive quantity.

\revision{Here we consider fermions on a lattice at finite temperature, so that}
the radius of convergence of the diagrammatic series 
is finite~\revision{\cite{Mastropietro,diagmchub,Rossi}}. Assuming that we are inside the radius of convergence, the convergence is exponential,
\begin{equation}
|a_n|  \underset{n \to \infty}{=} O( R^{-n})
\label{eq:an_Rn}
\end{equation}
with $R>1$ a constant.
Here and in what follows we omit 
multiplicative
constants and power laws which do not affect the dominant scaling behavior.


The number of diagrams scales factorially with the order $n$. For CDet, however, one takes into account cancellations between different diagram topologies. More specifically, we expect that
for fermions on a lattice at finite temperature,
\begin{eqnarray}
\!\!\!\!\!\!\!\!\!\!\!\!\!\!\!\!\!\! z_n &\underset{n\to \infty}{\sim}& 
                  R_D^{-n} ~n! \quad\quad\quad\quad\quad\quad\quad    {\rm (DiagMC)} \label{eq:RD} \\
  z_n &\underset{n\to \infty}{\sim}& R_C^{-n}   \quad \quad \quad\quad \quad  \quad \quad \quad \quad  {\rm (CDet)} 
               \label{eq:RTD}
              \end{eqnarray}
with $R_D$ and $R_C$ positive constants.\footnote{
\revision{ Equation~(\ref{eq:RTD}) is a natural conjecture given Eq.~(\ref{eq:an_Rn}); its}
rigorous proof
 may be obtained using techniques similar to those of
 Ref.~\cite{abdes} (J. Magnen, private communication).
\revision{Equation~(\ref{eq:RD}) is plausible since this is the generic large-order behavior for bosonic theories~\cite{ZinnRevue}.}
}


Let us discuss the behavor of the average sign,
$\langle{\rm sign}\rangle := \langle {\rm sign}\, A_n\rangle_{P_n}=a_n/z_n $,
as a function of the order $n$.
For DiagMC, we see that $\langle{\rm sign}\rangle$ tends to zero
 factorially, a manifestation of the near-compensation between different diagrams.
For CDet, $\langle{\rm sign}\rangle$ tends to zero exponentially
in the generic case where $R_C<R$.

As a result, the variance on individual measurements behaves as
\begin{eqnarray}
\sigma_{A_n} = z_n \sqrt{1 - \langle  {\rm sign}  \rangle^2_{P_n}}  \underset{n\to \infty}{\sim} z_n   \; ,
\label{eq:sigmaAn}
\end{eqnarray}
which together with 
Eq.~(\ref{eq:epsstat}) gives for the statistical error bar on the $n$-th order contribution $a_n$,
\begin{equation}
  \epsilon_{{\rm stat}}(n) \underset{n\to\infty}{\sim}  z_n\, \sqrt{\frac{2~\tau_{\rm auto}(n)+1}{N_{MC}(n)}}.
  \label{eq:estatN}
\end{equation}
Here $\tau_{\rm auto}(n)$ is expected to increase at most polynomially with $n$, 
which 
we have checked
numerically for CDet;
we will neglect this $n$-dependence of $\tau_{\rm auto}$  since it will not affect the final scalings.
An appropriate dependence $N_{MC}(n)$ of the number of MC steps on order will be specified below.

Note that the relative statistical error
$\epsilon_{\rm stat}(n)/a_n \propto z_n / a_n = 1/\langle{\rm sign}\rangle_{P_n}$ diverges factorially for DiagMC and exponentially for CDet.
This can be viewed as
 a  sign problem for diagrammatic Monte Carlo methods,
limiting the order that can be reached.
On the other hand, the exponential convergence (\ref{eq:an_Rn}), which is only possible thanks to the fermionic sign,
implies that reaching very high orders is not necessary, as we now quantify.

The assumption of
exponential convergence 
implies that the systematic error due to the finite diagram-order cut-off~$N$,
\begin{equation}
\epsilon_{\rm sys}(N) = \sum_{n=N+1}^{\infty} a_n,
\end{equation}
decreases exponentially,
\begin{equation}
\epsilon_{\rm sys}(N)     \underset{N \to \infty}{=}  O(R^{-N}) \; .
\label{eq:esysN}
\end{equation}
To achieve a final error $\sim\epsilon$, 
it is then natural to
work in a regime where systematic and statistical errors are on the same order.
We thus choose $N$ such that 
$R^{-N} \sim \epsilon$,
and we take a computational time $t$ such that the total statistical error is $\epsilon_{\rm stat}\sim\epsilon$.
Neglecting correlations between different orders,
we have $\epsilon_{\rm stat}^2 \simeq \sum_{n=0}^N \epsilon_{\rm stat}(n)^2$, which leads us to choose $N_{MC}(n)$ such that $\epsilon_{\rm stat}(n)$ is $n$-independent.
Equation~(\ref{eq:estatN}) together with Eqs.~(\ref{eq:RD},\ref{eq:RTD}) then yield
\begin{equation}
 t_n \underset{n\to\infty}{\sim}\left\{
                \begin{array}{ll}
\displaystyle
\frac{1}{\epsilon^2}\,\frac{(n!)^2}{R_D^{2n}}
                   \quad &   {\rm (DiagMC)} \\
\displaystyle
\frac{1}{\epsilon^2}\,\left(\frac{3}{R_C^2}\right)^n
                   \quad &  {\rm (CDet)} \; ,
                \end{array}
              \right.
\end{equation}
where the factor $3$ for CDet comes from the fact that the computational time per MC-step is $\sim 3^n$, because of the recursive formula that needs to be evaluated in order to eliminate disconnected diagrams~\cite{Rossi}.
As a result, for DiagMC most time is spent sampling the highest order, 
while for CDet this is the case only for $R_C<\sqrt{3}$.
Finally, we get
\begin{eqnarray}
 t(\epsilon) &\sim& \epsilon^{- \#  \ln (\ln \epsilon^{-1})}  \quad\quad\quad\quad\quad\quad    {\rm (DiagMC)}  \\
       t(\epsilon) &\sim&           \epsilon^{-\alpha}  \quad\quad\quad\quad\quad\quad\quad\quad\quad\quad {\rm  (CDet)} 
\label{eq:CCP_CDet}
\end{eqnarray}
Hence polynomial scaling is nearly reached with DiagMC,
and is achieved with CDet.
The exponent for CDet is given by
\begin{equation}
\alpha   =   2 +  \frac{ \ln(3/{R_C}^2) }{\ln R} 
\end{equation}
if $R_C < \sqrt{3}$.

The case $R_C > \sqrt{3}$ is particularly instructive. Here, most time is spent sampling low diagram orders, and one has $\alpha=2$, which is the best scaling one can achieve in any Monte Carlo computation. We thus conclude that fermionic sign---all by itself---does not necessarily leads to {\it any} qualitative effect on the scaling of computational time with $\epsilon$.

\begin{figure}[t]
\includegraphics[width=\columnwidth]{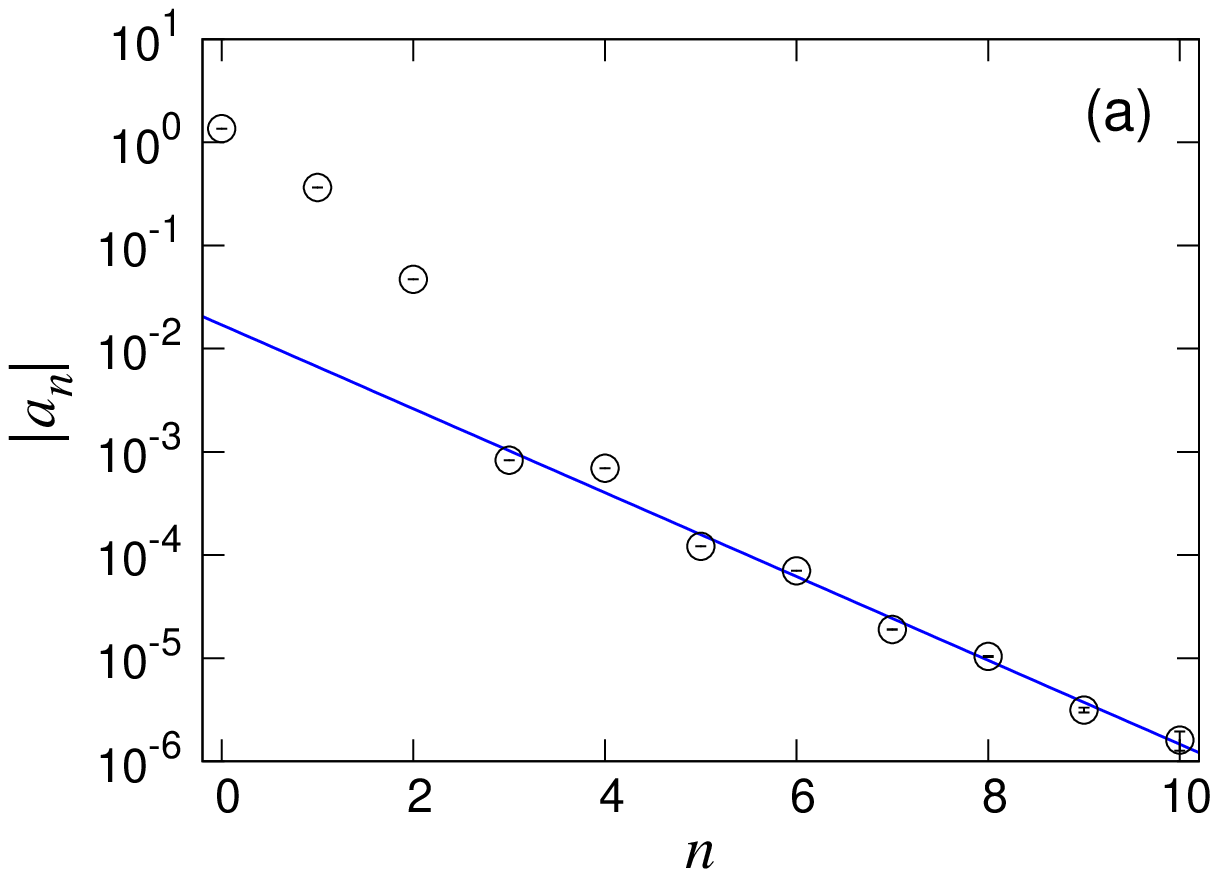}
\includegraphics[width=\columnwidth]{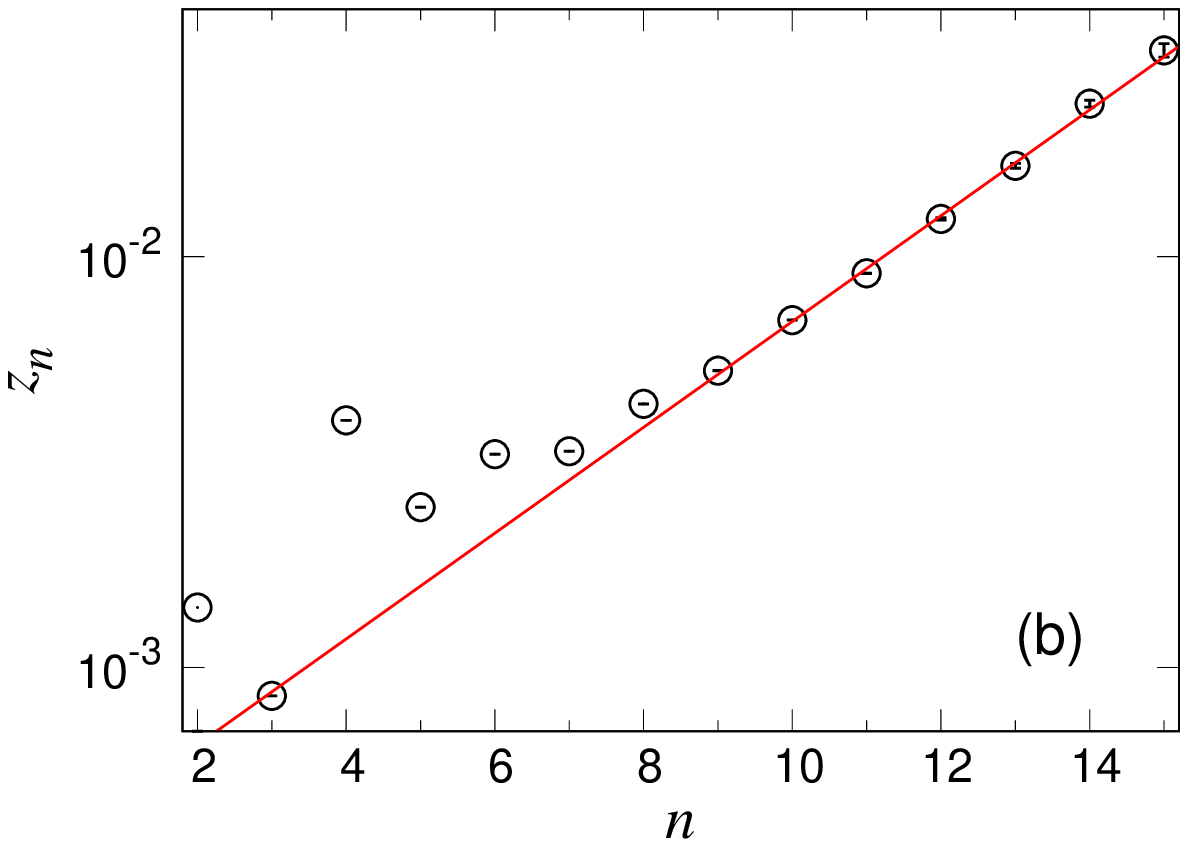}
\caption{(a)
Absolute value of the sum of all order-$n$ diagrams $|a_n|$, 
and (b)
 weight $z_n$ 
of the order-$n$ configuration space of Connected Determinant Diagrammatic Monte Carlo, 
for the pressure of the Fermi-Hubbard model
(at $U=2$, $\beta=8$, $n \simeq 0.875$).
The lines are linear fits to the data at large $n$.
\label{Fig1}}
\end{figure}

 The above scalings
are not purely academic considerations,
as we illustrate with an example for CDet.
We analyse the computation,
reported
in Ref.~\cite{Rossi},
of the pressure of Fermi-Hubbard model in two dimensions.
The diagrammatic scheme is a bare series
with bare tadpoles taken into account through a shift
 of the chemical potential. 
The
Hubbard parameters are: interaction $U = 2$, chemical potential $\mu=0.55978$ and inverse temperature $\beta=8$ (with hopping $=1$);
this corresponds to a  density $n = 0.87500(2)$.
Figure~\ref{Fig1}(a) shows
 that $|a_n|$ approaches an exponential behavior $R^{-n}$
with $R=2.5(1)$, 
while Figure~\ref{Fig1}(b) shows that $z_n$ approaches $R_C^{-n}$
with $R_C=0.75(3)$.\footnote{For the density and the kinetic energy, we find the same value of $R_C$ within our error bars.}
We can make three important observations.
First,
the exponent
$\alpha=3.8(2)$
is not too large.
Second, we clearly reach the asymptotic regime where Eqs.~(\ref{eq:RTD},\ref{eq:esysN}), and therefore also Eq.~(\ref{eq:CCP_CDet}), are valid.
Third, 
$|a_n|$ at low $n$ is ${\sim} 100$ times
larger
than the extrapolation
to low orders of the large-order behavior shown by the straight line in Fig.~1(a).
These three observations
explain why it was possible to obtain a ${\sim}10^{-6}$ relative accuracy for the pressure in Ref.~\cite{Rossi}.
Interestingly, the second and third observations hold independently of $U$ (with $\mu(U)=\mu_0+U n_0/2$ as in~\cite{Rossi}).
In contrast, $\alpha$ diverges when $U$ tends to the critical value $U_c = 2\,R(U=2)\simeq 5.1$ such that $R(U_c)=1$.
Divergent-series summation methods may allow to approach this point and even to go beyond it; we leave this for future study.

To avoid possible confusion, 
let us remark that 
we do not claim any connection between our results and
the computational complexity theory of computer science.
In this theory, a `problem instance' is defined by $\mathcal{N}$ parameters,
and
the P complexity-class is defined by polynomial scaling of computational time {\it with respect to $\mathcal{N}$ for $\mathcal{N}\to\infty$} (in the worst case with respect to all possible instances).
In the spin-glass problem 
discussed
 in~\cite{TroyerWieseNP},
an instance is a disorder realisation, and
$\mathcal{N}$ is the number of random couplings (which happens to coincide with the system volume).
In contrast, 
here we consider 
problems 
defined by a {\it fixed} (usually small) number of model-parameters (e.g. $T$, $U$ and $\mu$ for the Hubbard model in the thermodynamic limit).

We turn to `traditional' QMC, by which we mean here path-integral or auxiliary-field QMC.
We consider the computation of an intensive quantity at finite temperature.
Generically the FSP leads to an exponential scaling 
with spatial volume and inverse temperature
of the average sign,
and hence, due to Eq.~(\ref{eq:epsstat}), of the
 statistical error~(see, e.g., \cite{Loh,CeperleyPIQMC,AssadChapter,troyer_topol_origin}):
\begin{equation}
\epsilon_{\rm stat}(L) \underset{L\to\infty}{\sim} \frac{e^{\# \beta L^d}}{\sqrt{t}} \; ,
\label{eq:errL}
\end{equation}
where $L$ is the linear system size. 

Beside the statistical error, we also need to take into account the systematic error $\epsilon_{\rm sys}(L)$ coming from the finite size $L$.
The total error $\epsilon$ entering the CCP is $\epsilon \sim \epsilon_{\rm stat}(L)+\epsilon_{\rm sys}(L)$.
We assume that finite-size corrections decrease exponentially,
\begin{equation}
 \epsilon_{\rm sys}(L) \underset{ L \to \infty}{\sim} 
                  e^{-\# L},
              \label{eq:Lerr}
              \end{equation}
which is expected generically (away from second-order phase transitions and at finite temperature).
For a given computational-time $t$, the optimal strategy is to choose $L$ so that
$\epsilon_{\rm sys}\sim \epsilon_{\rm stat}$, which yields
\begin{equation}
 t(\epsilon) \sim
                  e^{ \# \beta(\ln \epsilon^{-1})^d}.
              \label{eq:ccppi}
               \end{equation}
So for $d>1$ the scaling of $t$ with $\epsilon^{-1}$ is quasi-polynomial and there is a CCP.
In one dimension there is no CCP, which is another illustration of the simple observation presented in the introduction.


Apart from these asymptotic scalings, there are also practical advantages of diagrammatic methods over traditional QMC.
For traditional QMC, the condition for getting close to the thermodynamic limit
 is typically $L$ much larger than the correlation length.
Equation (\ref{eq:errL}) then yields a computational time $t\propto e^{\# 2\beta L^d}$
which is often prohibitive, meaning that one cannot get close to the thermodynamic limit, and that
one cannot even reach
the asymptotic scaling regimes~(\ref{eq:Lerr},\ref{eq:ccppi}). 
The situation is very different in diagrammatic expansions, 
where as shown by the above example, the asymptotic regime is accessible, and moreover the lowest orders typically set the scale while higher-order
contributions are merely corrections.

In conclusion, \revision{unbiased} numerical methods for solving quantum many-fermion problems should be evaluated on the basis
of their scaling of computational time with respect to the final error bar on thermodynamic-limit quantities.
The presence of a fluctuating sign 
does not suffice to say that a problem is intractable by Monte Carlo.
Nothing prevents in principle a polynomial scaling of
the CPU-time versus  the inverse error bar.
We demonstrated that such polynomial complexity is indeed achieved by
the recently proposed CDet method when inside the radius of convergence of the Feynman diagrammatic series.
Since this method offers the possibility to calculate properties of many-fermion systems in polynomial time, it is fair to say that the sign problem has become irrelevant here and a numerical
solution to the many-fermion problem is available, at least in some region of parameter space.

\acknowledgements
We acknowledge support from ERC Grant {\it Thermodynamix} (NP and FW), the Simons Foundation's {\it Many Electron Collaboration},
the National Science Foundation under the grant PHY-1314735, and the MURI Program {\it New Quantum Phases of Matter} from AFOSR
 (NP and BS). 
Some of us are members of {\it Paris Sciences et Lettres}, {\it Sorbonne Universit\'es} (RR, KVH and FW) and {\it Sorbonne Paris-Cit\'e} (RR and KVH).


\begin{thebibliography}{99}


\bibitem{Loh} E. Y. Loh, J. E. Gubernatis, R. T. Scalettar, S. R. White, D. J. Scalapino, R. L. Sugar,
Phys. Rev. B {\bf 41}, 9301 (1990).

\bibitem{CeperleyPIQMC} D. M. Ceperley, {\it Path integral Monte Carlo methods for fermions}, in {\it Monte Carlo and Molecular Dynamics of Condensed Matter Systems}, Ed. K. Binder and G. Ciccotti, Editrice Compositori, Bologna, Italy (1996).

\bibitem{CeperleyHouches} D. M. Ceperley, {\it Quantum Monte Carlo Methods for Fermions}, Proceedings of the Les Houches Summer School, Session 56, {\it Strongly Interacting Fermions and High $T_{c}$ Superconductivity}, eds. B. Doucot and J. Zinn-Justin, Elsevier (1995).

\bibitem{AssadChapter}
F.F. Assaad and H.G. Evertz,
{\it World-line and Determinantal Quantum Monte Carlo Methods for Spins, Phonons and Electrons},
in {\it Computational Many-Particle Physics},
H. Fehske, R. Schneider and A. Wei{\ss}e (Eds.), Springer,
Lect. Notes Phys. {\bf 739}, 277-356 (2008).




\bibitem{WieseMeron}
S. Chandrasekharan
and U.-J. Wiese, Phys. Rev. Lett.
{\bf 83}, 3116 (1999).

\bibitem{SachdevSignFree}
E. Berg, M. A. Metlitski, and S. Sachdev, Science {\bf 338}, 1606
(2012).

\bibitem{Wei}  Z. C. Wei, C. Wu, Yi Li, S. Zhang, and T. Xiang, Phys. Rev. Lett. {\bf 116}, 250601 (2016).

\bibitem{AletSignFreeFrustratedSpins}
F. Alet, K. Damle, S. Pujari,
Phys. Rev. Lett. {\bf 117}, 197203 (2016).

\bibitem{HoneckerSignFreeFrustratedSpins}
A. Honecker, S. Wessel, R. Kerkdyk, T. Pruschke, F. Mila, B. Normand,
Phys. Rev. B {\bf 93}, 054408 (2016)


\bibitem{troyer_topol_origin}
M. Iazzi, A. A. Soluyanov, and M. Troyer,
Phys. Rev. B
{\bf 93}, 115102 (2016).

\bibitem{capponi_honeycomb}
S. Capponi, J. Phys.: Condens. Matter {\bf 29}, 043002  (2017).

\bibitem{Rossi} R. Rossi, arxiv:1612.05184.


\bibitem{diagmchub} K. Van~Houcke, E.~Kozik, N.~Prokof'ev and B. Svistunov, {\it Diagrammatic Monte Carlo}, in ``Computer Simulation Studies in Condensed Matter Physics XXI. CSP-2008''  (Eds.
D.P. Landau, S.P. Lewis, and H.B. Sch{\"u}ttler), Physics Procedia {\bf 6}, 95 (2010).




\bibitem{Mastropietro} \revision{G. Benfatto, A. Giuliani, and V. Mastropietro, Annales H. Poincar{\'e} {\bf 7}, 809 (2006).}

\bibitem{abdes} A. Abdesselam and V. Rivasseau, Lett. Math. Phys. {\bf 44}, 77 (1998).

\bibitem{ZinnRevue} \revision{J. Zinn-Justin, Phys. Rep. {\bf 70}, 109 (1981).}



\bibitem{TroyerWieseNP}
M. Troyer and U.-J. Wiese,
Phys. Rev. Lett. {\bf 94}, 170201  (2005).

\end{thebibliography}
\end{document}